\RequirePackage{fix-cm}
\documentclass[smallextended,natbib]{svjour3}       
\smartqed  
\usepackage{graphicx}
\usepackage{amsmath}
\usepackage{amsfonts}
\newcommand{\vv}[1]{\mbox{\boldmath $#1$}}
\newcommand{\mtr}[1]{\mathbf{#1}}
\newcommand{\iu}{\mathrm{i}}
\newcommand{\qq}[1]{\mathsf{#1}}
\newcommand{\rd}{\mathrm{d}}
\newcommand{\cH}{\mathcal{H}}
\newcommand{\cK}{\mathcal{K}}
\newcommand{\pb}[2]{\left\{ #1, #2\right\}}

\journalname{CELE}
\begin{document}

\title{Kustaanheimo-Stiefel transformation with an arbitrary defining vector}

\titlerunning{KS transformation with an arbitrary defining vector}        

\author{S. Breiter         \and
        K. Langner
}


\institute{S. Breiter \at
              Astronomical Observatory Institute, Faculty of Physics, Adam Mickiewicz University, Sloneczna 36, 60-286 Poznan, Poland \\
              \email{breiter@amu.edu.pl}           
           \and
           K. Langner \at
             Astronomical Observatory Institute, Faculty of Physics, Adam Mickiewicz University, Sloneczna 36, 60-286 Poznan, Poland \\
              \email{krzysztof.langner@amu.edu.pl}           
}

\date{Received: 7 November 2016 / Accepted: date}

\maketitle

\begin{abstract}
Kustaanheimo-Stiefel (KS) transformation depends on the choice of some preferred direction in the Cartesian 3D space.
This choice, seldom explicitly mentioned, amounts typically to the direction of the first
or the third coordinate axis in celestial mechanics and atomic physics, respectively.
The present work develops a canonical KS transformation with an arbitrary preferred direction, indicated by what we call
a defining vector. Using a mix of vector and quaternion algebra, we formulate the transformation in a reference frame independent manner.
The link between the oscillator and Keplerian first integrals is given. As an example of the present formulation,
the Keplerian motion in a rotating frame is re-investigated.

\end{abstract}

\section{Introduction}
\label{intro}

There are many ways to convert the Kepler problem into the isotropic harmonic oscillator. A comprehensive summary
can be found in the monograph by \citet{Cordani} and in the work of \citet{DEF}, the two sources overlapping  only partially.
Among all the methods, the Kustaanheimo-Stiefel (KS) transformation stands exceptional by its simplicity, popularity, and history
(traced back to Heinz Hopf, or even Carl Friedrich Gauss -- see \citet{Volk:76}). The literature concerning
theoretical and practical aspects of the KS variables is already vast, yet there is a feature which, to our knowledge, has not
brought enough attention by either being too obvious, or by not being realized. We mean the existence of some preferred direction
in the definition of the KS variables. Kustaanheimo, Stiefel, and most of their followers in the realm of celestial mechanics,
use the set of variables so designed, that only the first Cartesian coordinate $x_1$ involves the squares of the KS coordinates.
In the present paper, we will refer to it as KS1. However, in atomic physics a different set, to be named KS3, is considered standard,
at least since late 1970s \citep[e.g.][]{DuKl:79}. There, only the $x_3$ involves the squares.
The choice of the preferred direction is of marginal significance for the unperturbed Keplerian problem,
but it may either simplify, or complicate expressions resulting from added perturbations. Thus we have found it worthwhile
to establish a general KS transformation with  the preferred direction left unspecified.

In section~\ref{sec:1} we give a brief outline of how the KS1 transformation has settled down in the realm of celestial mechanics.
Not attempting a detailed bibliographic survey, we mark five turning points: i) the first paper of \citet{Kust:64} where the
transformation is born in spinor form, ii) the metamorphosis into the L-matrix setup done by \citet{KS:65}, iii)
early quaternion formulation by \citet{Viv:83}, iv) the refinement of the quaternionic form due to \citet{DEF},
and v) interpretation of the KS variables as rotation parameters \citep{Saha:09}. Save for the last point, the reader may observe how
the special role of the $Ox_1$ axis direction is transferred from one setup to another, and how the presence of a preferred direction
becomes more and more evident.

Once the stage has been set, we discuss the main theme in Section~\ref{KSarb}. Introducing the notion of a defining vector,
we build the KS transformation (and its canonical extension) with an arbitrary preferred direction. Working within the general quaternion
and vector formalism, we refrain from using explicit expressions in terms of coordinates. Hopefully, it should allow capturing
the intrinsic, coordinate independent features of the KS transformation.

Section~\ref{hamil} considers the general properties of the Hamiltonian in KS variables.
Special attention is paid to linking the invariants of the unperturbed problem in its two incarnations: the Kepler problem in
Cartesian coordinates, and the isotropic oscillator in KS variables.

Finally, in order to give an example that the choice of the defining vector does matter in a perturbed problem, we
return to the Kepler problem in the rotating reference frame. Compared to earlier works which used the KS1 set, the task
can be considerably facilitated by the appropriate selection of the preferred direction, as demonstrated in section~\ref{keprot}.

\section{KS1 transformation}
\label{sec:1}
\subsection{The roots}
\label{roots}
What Paul Kustaanheimo announced at the Oberwolfach conference on Mathematical Methods of Celestial Mechanics and published
the same year \citep{Kust:64} is worth a brief recall, because neither \textit{Annales Universitatis Turkuensis}, nor the
\textit{Publications of the Astronomical Observatory of Helsinki} are widespread enough. Moreover, an awkward notation
has masked some features that emerge immediately when the more common conventions are applied.

Given a Cartesian position vector $\vv{x}=(x_1,x_2,x_3)^\mathrm{T}$,
Kustannheimo extended it formally to a null 4-vector of the Minkowski space, using  the length $r = \sqrt{\vv{x}\cdot \vv{x}} = x_0$
as an extra coordinate. These served as the coefficients
 for a linear combination of a unit matrix ${\mtr{\sigma}}_0$ and the Pauli matrices
\citep{Cartan}
\begin{equation} \label{Pauli}
 \sigma_1=\left(\begin{array}{rr} 0 & 1\\ 1 & 0 \end{array}\right),\quad \sigma_2=\left(\begin{array}{rr} 0 & - \iu \\ \iu & 0 \end{array}\right),
\quad \sigma_3=\left(\begin{array}{rr} 1 & 0\\ 0 & -1\end{array}\right).
\end{equation}
The result is a complex matrix
\begin{equation} \label{Xmat}
\mtr{S}= r \sigma_0  +  x_1 \sigma_{3} + x_2 \sigma_{1} - x_3  \sigma_{2}=
\left(\begin{array}{cc} r+x_1 & x_2 + \iu \, x_3 \\ x_2-\iu \,  x_3 & r-x_1 \end{array}\right),
\end{equation}
assigned to the vector $(x_0,x_1,x_2,x_3)^\mathrm{T}$.
The difference between the treatment of $x_1$ in the diagonal and the complex pair $x_2 \pm \iu x_3$
can be spotted already at this stage.\footnote{The disparity of indices associated with $x_i$ and $\sigma_j$,
was originally not visible, since Kustaanheimo
used a different set of the Pauli matrices (namely: $\mtr{i}_x = \sigma_3$, $\mtr{i}_y = \sigma_1$, and $\mtr{i}_z = - \sigma_2$).
It took some time until the physicists swapped to $\mtr{S}= \sum_{j=0}^3 x_j \sigma_j$, the KS3 convention.}

Any Hermitian matrix $\mtr{S}$ can be expressed in terms of two complex numbers $S_1$ and $S_2$. Using
a complex 2-vector $\vv{s} = (S_1,S_2)^\mathrm{T}$, \citet{Kust:64} defined $\mtr{S}$ in terms of its Hermitian outer product
\begin{equation} \label{Smat}
\mtr{S}= 2  \left(
                     \begin{array}{c}
                       S_1 \\
                       S_2 \\
                     \end{array}
                   \right) \left(
             \begin{array}{cc}
               \overline{S}_1 & \overline{S}_2 \\
             \end{array}
           \right) =
2\left(\begin{array}{cc} S_1 \overline{S}_1 & S_1 \overline{S}_2 \\ \overline{S}_1 S_2  & S_2 \overline{S}_2 \end{array}\right).
\end{equation}
By the equivalence of (\ref{Xmat}) and (\ref{Smat}), the vector $\vv{s}$ becomes a rank 1 spinor,\footnote{
Intriguingly, although the initial name of the KS transformation was `the spinor regularization', \citet{Kust:64} uses the word `spinor'
only twice: once in the title, and once in the abstract. \citet{KS:65} use it only once -- in the title.}
and the matrix $\mtr{S}$ -- a rank 2 spinor associated with $\vv{x}$ \citep[c.f.][]{BFM:87,Steane:13}.
Equating respective elements of (\ref{Xmat}) and (\ref{Smat}), one readily finds
\begin{eqnarray}
x_1&=& S_1 \overline{S}_1-S_2 \overline{S}_2, \nonumber \\
x_2&=& \overline{S}_1 S_2+S_1 \overline{S}_2, \label{Ktr}\\
x_3&=& \iu \, \left( \overline{S}_1 S_2 -S_1 \overline{S}_2\right), \nonumber \\
 r & = & S_1 \overline{S}_1+S_2 \overline{S}_2. \nonumber
\end{eqnarray}
As noted by Kustaanheimo, the transformation is not unique; indeed it involves only the products with conjugates,
so using any spinor $\vv{q} = (Q_1, Q_2)^\mathrm{T} = \vv{s} \exp{\iu \phi}$, leads to the same result. Geometrically,
it means that any pair of complex numbers resulting from rotations of $S_1$ and $S_2$ on a complex plane
by the same angle, generates the same position vector $\vv{x}$ in (\ref{Ktr})

The regularization of the Keplerian motion does not appear until the coordinate transformation (\ref{Ktr})
is augmented by the time transformation of the Sundman type. There, Kustaanheimo proposed a general formula relating the pseudo-time
$\tau$ to its physical counterpart $t$
\begin{equation}\label{sund-K}
    \frac{\rd \tau}{\rd t} = \frac{\beta}{r}\,\exp{\int K \, \rd t},
\end{equation}
where $K$ could be an arbitrary function of position, velocity and time. This flexibility has never been seriously explored,
and the simplest choice of $K=0$ has became standard.
The regularization converts the Kepler problem with energy constant $h$ into a spinor oscillator problem
\begin{equation}\label{spKep}
    \vv{s}'' =  \frac{h}{2} \vv{s},
\end{equation}
yet, this simple form does not show up, until the bilinear constraint
\begin{equation}\label{ivmS}
    \overline{S}_1 S'_1 - \overline{S}'_1 S_1 +  \overline{S}_2 S'_2 - \overline{S}'_2 S_2 = 0,
\end{equation}
is imposed,\footnote{Another wording of condition (\ref{ivmS}), provided by \citet{Kust:64},
is: $\overline{S}_1 S'_1 + \overline{S}_2 S'_2$, being a half of $r'$, is a real quantity.}
where the prime stands for the derivative with respect to the Sundman time $\tau$. Kustaanheimo justified
this choice by observing the invariance of the left hand side of (\ref{ivmS})  in the perturbed Kepler problem
with a particular form of perturbation (linear in coordinates, velocities and angular momentum).

One can only speculate what would be the fate of the Kustaanheimo's transformation, has it not attracted
the attention of Eduard Stiefel who coauthored the paper published next year \citep{KS:65}. In the new mise-en-sc\'ene,
the complex variables and unnecessary generalization were dropped, and the discussion focused on a real 4-vector
$\vv{u}=(u_1,u_2,u_3,u_4)^\mathrm{T}$, containing the parameters of the substitution
\begin{equation}\label{S1S2}
    S_1 = u_1 + \iu \, u_4, \qquad S_2=u_2 - \iu \, u_3.
\end{equation}
Then, eq. (\ref{Ktr}) takes the form
\begin{eqnarray}
x_1 &=& u_1^2-u_2^2-u_3^2+u_4^2, \nonumber \\
x_2 &=& 2(u_1 u_2 - u_3 u_4), \label{KS}\\
x_3 &=& 2(u_1 u_3 + u_2 u_4), \nonumber \\
r & = & u_1^2+u_2^2+u_3^2+u_4^2, \nonumber
\end{eqnarray}
and the constraint (\ref{ivmS}) is turned into
\begin{equation}\label{inv2}
     u_4 u_1' - u_3 u_2' + u_2 u_3' - u_1 u_4' = 0.
\end{equation}
This derivation, however, cannot be found in the paper; \citet{KS:65} have burnt the bridge leading back to
the 1964 work and started the presentation from the matrix equation that related
$\vv{u}$ with a 4-vector $\vv{x} =(x_1,x_2,x_3,0)^\mathrm{T}$
through the  matrix product
\begin{equation}\label{KSvec}
    \vv{x}  = \mtr{L}(\vv{u})\,\vv{u}.
\end{equation}
The L-matrix definition, given by \citet{KS:65} with an intriguing clause `for example', was
\begin{equation}
\mtr{L}(\vv{u})=\left(\begin{array}{rrrr}
  u_1 &-u_2& -u_3 & u_4
\\u_2 & u_1& -u_4 & -u_3
\\u_3 & u_4 & u_1 &  u_2
\\u_4& -u_3& u_2 &  -u_1 \end{array}\right). \label{matrixL}
\end{equation}
This definition leads to the first three equations (\ref{KS}) for $x_j$, whereas the last of equations (\ref{KS})
had been postulated as a required property of $\mtr{L}(\vv{u})$. Noteworthy, the special role of $x_1$
has been conserved, as visible in the first of equations (\ref{KS}): other coordinates are defined by products
$u_i u_j$, whereas $x_1$ is composed of the pure squares $u_i^2$.

The work, published in a more widespread journal and written in a manner friendly to
the celestial mechanics audience, considerably helped to promulgate what is now known as the Kustaanheimo-Stiefel transformation.
By the influence of the \citet{StS:71} monograph, the matrix approach became paradigmatic in the celestial mechanics community,
and the `for example' choice (\ref{matrixL}) has been taken for granted and obvious, save for occasional renumbering of indices
and the change of sign in $u_4$.

\subsection{Enter quaternions }

The close relation between spinors and quaternions was known already to Cartan. But in the framework of the L-matrix formulation,
the relation of the KS variables to the quaternion algebra is merely an additional aspect,
mentioned by \citet{KS:65} or \citet{StS:71} as an interesting, but probably unimportant curio. Setting the KS transformation
in the quaternion formalism, originated by \citet{Viv:83} and applauded by \citet{DEF}, offered new paths  to understanding
the known properties of the transformation, as well as to its generalization to higher dimensions.

Following \citet{DEF} we treat a quaternion $\qq{v} = (v_0,v_1,v_2,v_3)$, or $\qq{v}=(v_0,\vv{v})$, as a union of a scalar $v_0$ and of
a vector $\vv{v} = (v_1,v_2,v_3)^\mathrm{T}$.
Conjugating a quaternion, we change the signs of its vector part, i.e.
\begin{equation}\label{conq}
    \bar{\qq{v}} = (v_0,-v_1,-v_2,-v_3) = (v_0 , - \vv{v}).
\end{equation}
Extracting the vector part is performed by means of the operator $\natural$, so that for $\qq{v}=(v_0,\vv{v})$
\begin{equation}\label{nat}
 \vv{v} = \qq{v}^\natural.
\end{equation}

The usual scalar product, marked with a dot,
\begin{equation}\label{scpq}
    \qq{v} \cdot \qq{w} = v_0 w_0 + \vv{v} \cdot \vv{w},
\end{equation}
is commutative, but the quaternion product
\begin{equation}\label{qp}
  \qq{v}\, \qq{w} = \left( v_0 w_0 - \vv{v} \cdot \vv{w} , v_0 \vv{w} + w_0 \vv{v} +  \vv{v} \times \vv{w}\right),
\end{equation}
is not.
Using a standard basis
\begin{equation}\label{baza}
    \qq{e}_0 = (1,\vv{0}), \quad  \qq{e}_1 = (0,\vv{e}_1),  \quad \qq{e}_2 = (0,\vv{e}_2), \quad \qq{e}_3 = (0,\vv{e}_3),
\end{equation}
we recover the classical `$1ijk$' multiplication rules of Hamilton for the basis quaternions: $\qq{e}_0 \qq{e}_1 = \qq{e}_1$
for $1 i = i$,
$\qq{e}_1 \qq{e}_2 = \qq{e}_3$ for $ij=k$, etc.
The inverse of a quaternion is, in full analogy with complex numbers,  $\qq{v}^{-1} = \bar{\qq{v}}/|\qq{v}|^2$,
where the norm is $|\qq{v}| = \sqrt{\qq{v}\cdot \qq{v}}$.
The conjugate of a product is, typically for noncommutative operations,
\begin{equation}\label{coru}
    \overline{\qq{u} \qq{v}} = \bar{\qq{v}} \bar{\qq{u}}.
\end{equation}
Let us observe a useful property of the mixed dot product,
\begin{equation}\label{dper}
     \bar{\qq{u}} \cdot ( \qq{v}   \qq{w} ) = \bar{\qq{w}} \cdot ( \qq{u} \qq{v}) = \bar{\qq{v}} \cdot (\qq{w} \qq{u}),
\end{equation}
echoing the mixed product rule of the standard vector algebra.

Another useful operation is called a quaternion outer product \citep{MGS:14} or a quaternion cross product \citep{StS:71,Viv:88,DEF}.
Conventions vary among the authors; we adopt the one of \citet{DEF}
\begin{equation} \label{qop}
\qq{u} \wedge \qq{v} = \frac{ \qq{v} \bar{\qq{u}} - \qq{u} \bar{\qq{v}} }{2}
= \left( 0, \, u_0 \vv{v} - v_0 \vv{u} + \vv{u} \times \vv{v} \right).
\end{equation}
A remarkable property of the cross product, not mentioned by \citet{DEF}, is a factor exchange rule
\begin{equation}\label{cper}
    (\qq{u} \qq{v}) \wedge \qq{w} = \qq{u} \wedge (\qq{w} \bar{\qq{v}}),
\end{equation}
following directly from (\ref{qop}) and (\ref{coru}).

Note that the cross product always results in a quaternion with null scalar part, called a pure vector \citep{DEF} or, more often,
a pure quaternion \citep[e.g.][]{MGS:14}. We adopt the former convention.

Casting the KS transformation in a quaternion form is not a novelty. It can be found already in \citet{StS:71}.
In spite of the discouraging comments attached by the authors,
\citet{Viv:83} returned to this formalism and issued a different quaternion form of the transformation.
Feeling obliged to adhere to the `for example' convention of \citet{KS:65},
she reconstructed the transformation (\ref{KS}) as a quaternion product
\begin{equation}\label{KSV}
    (x_1,x_2,x_3,0) = \qq{u} \, \left[ \qq{e}_3 \, \bar{\qq{u}} \, \bar{\qq{e}}_3 \right] = \qq{u} \, \qq{u}_\ast,
\end{equation}
where $\qq{u} = (u_1,u_2,u_3,u_4)$, and the product in square brackets is an `anti-involute' $\qq{u}_\ast$ of $\qq{u}$
(an operation, that actually amounts to a trivial change of sign in $u_4$)\footnote{ The `anti-involute' was
later reinvented as a `star conjugate' by \citet{Wald:06,Wald:08}. His formulation of KS1,  similar to (\ref{KSV}),
offers an interesting interpretation of the invariant (\ref{inv:2}).} . Of course, the presence of $\qq{e}_3$
in (\ref{KSV}) does not mean that the direction of $x_3$ gains some special position; the transformation of Vivarelli
remains the pure KS1.

\citet{DEF}, decided to link a more natural, vector type  quaternion $\qq{x} = (0,x_1,x_2,x_3)$ with a KS quaternion
$\qq{v} = (v_0,v_1,v_2,v_3)$ and then, not needing an anti-involute, found that\footnote{setting  their additional parameter
$\alpha=1$.}
\begin{equation}\label{KSD}
     \qq{x} = \qq{v} \, \qq{e}_1 \, \bar{\qq{v}},
\end{equation}
which is not far from the original quaternion formulation of \citet{StS:71}.
Converting a Stiefel-Scheifele-Vivarelli quaternion $\qq{u}$ to a Deprit-Elipe-Ferrer quaternion $\qq{v}$
can be effected by the rule
\begin{equation} \label{DtoKS}
(v_0,v_1,v_2,v_3) \rightarrow (-u_4,u_1,u_2,u_3).
\end{equation}
With this rule, the outcome of (\ref{KSD}) is equivalent to the transformation of \citet{KS:65}. In particular,
the distinguished role of $x_1$ remains unaffected, and clearly marked by the presence of $\qq{e}_1$ in equation
(\ref{KSD}), as it will follow from the interpretation given below.

\subsection{Kustaanheimo-Stiefel meet Euler-Rodrigues}
\label{KSER}

For practitioners, the most enjoyed property of quaternions is their straightforward relation to rotation.
Rotation of a vector $\vv{x}$, formally treated as a quaternion with null scalar part $(0,\vv{x})$,
is specified by a unit quaternion $\qq{q}$, including the complete information about the rotation
angle $0 \leq \theta \leq \pi$, and rotation axis given by the unit vector $\vv{n}$. Then, with
\begin{equation}\label{qrot}
    \qq{q} = \left( \cos\frac{\theta}{2}, \sin\frac{\theta}{2} \, \vv{n} \right),
\end{equation}
the rotated vector $\vv{y}$ is obtained through the quaternion product
\begin{equation}\label{qr1}
    (0,\vv{y}) = \qq{q} \, (0,\vv{x}) \, \bar{\qq{q}}.
\end{equation}
If we write the unit quaternion appearing in (\ref{qr1}) simply as $\qq{q}=(q_0,q_1,q_2,q_3)$,
its components are the Euler-Rodrigues parameters (sometimes called the Cayley parameters) of
the rotation matrix. Indeed, skipping the scalar part, equation (\ref{qr1}) may be rewritten in the matrix form as
\begin{equation}\label{rot}
    \vv{y} = \mtr{R}(\qq{q}) \vv{x},
\end{equation}
with the rotation matrix
\begin{equation}\label{Rotm}
   \mtr{R}(\qq{q}) = \left(
               \begin{array}{rrr}
                 q_0^2 + q_1^2 - q_2^2 - q_3^2 & 2 \, ( q_1 q_2 - q_0 q_3 ) & 2\,( q_0 q_2 +  q_1 q_3) \\
                 2 \, (q_1 q_2 +  q_0 q_3 ) & \quad q_0^2 - q_1^2 + q_2^2 - q_3^2 & -2 \,( q_0 q_1 -  q_2 q_3) \\
                 -2 \, ( q_0 q_2 -  q_1 q_3 ) & 2 \, ( q_0 q_1 + q_2 q_3) & \quad q_0^2 - q_1^2 - q_2^2 + q_3^2 \\
               \end{array}
             \right).
\end{equation}

This property reveals the meaning of the quaternion KS transformation (\ref{KSD}):\\
\begin{center}
\framebox{%
\begin{minipage}{\textwidth}
\begin{quote}
    up to the reassignment (\ref{DtoKS}), the normalized Kustaanheimo-Stiefel variables $u_i/\sqrt{r}$ are the Euler-Rodrigues
    parameters of the rotation turning the unit vector of the first axis $\vv{e}_1$ into the unit radius vector $\vv{x}/r$.
\end{quote}
\end{minipage}
}
\end{center}

Thus we find another argument in favor of the claim that the KS1 transformation attaches a special role
to the axis $Ox_1$.

The existence of some relation between KS variables and rotation was mentioned `for the record' by \citet{StS:71},
who declared the lack of interest in studying it closer. Then \citet{Viv:83} returned to this issue, but her description
is based on a statement that since a unit quaternion $\qq{q}$ `represents a rotation', so a product $\qq{q} \qq{q}_\ast$
also `represents' some rotation with the axis and angle expressions provided. But, unlike (\ref{KSD}),
the assignment $\qq{x} = \qq{q} \qq{q}_\ast$ is not a formula for rotation of some vector, leaving the whole argument in suspense.
It took years until \citet{Saha:09}, issued an explicit reference to the rotation.
His variant of the KS transformation
\begin{equation}\label{KSS}
    (0,\vv{x}) = \bar{\qq{v}} \, \qq{e}_3 \, \qq{v},
\end{equation}
differs from (\ref{KSD}) in two aspects: the conjugation sequence is different, and the basis quaternion $\qq{e}_3$ is used
instead of $\qq{e}_1$ (a rare example of the KS3 convention in celestial mechanics). According to \citet{Saha:09},
equation (\ref{KSS}) implies the rotation of the third axis to the $\vv{x}$ direction, although actually it describes the inverse rotation:
$\vv{x}$ to $\vv{e}_3$. Then -- up to the signs mismatch -- the right-hand side of (\ref{KSS}) coincides with the third column
of the matrix $\mtr{R}(\vv{v})$.

\section{KS transformation with arbitrary defining vector}
\label{KSarb}

\subsection{Point transformation}

The most straightforward generalization of the quaternion formulation proposed by \citet{DEF} is to consider
an arbitrary unit quaternion
\begin{equation}
\qq{c} = \sum_{j=0}^3  c_j \qq{e}_j,
\end{equation}
and the transformation
\begin{equation}\label{KSgen}
   \alpha \qq{x}  = \qq{v} \, \qq{c} \, \bar{\qq{v}},
\end{equation}
where a positive real parameter $\alpha$, having the dimension of length, is introduced
as in \citet{DEF} to allow the components of $\qq{v}$ have the same dimension as $\qq{x}$, as well
as to facilitate a comfortable units choice later on.
Since the scalar component of (\ref{KSgen}) is
$x_0 = c_0 \, \qq{v} \cdot \qq{v}$, and $c_0$ does not appear
in the vector part of $\qq{x}$, we may simply set $c_0=0$. Thus $\qq{c} = (0,\vv{c})$, and its
vector part will be called a defining vector. By definition, we assume
the unit length of the defining vector $||\vv{c}||=1$ throughout the text.

The remaining subsystem of (\ref{KSgen}) may be set in the matrix-vector form
\begin{equation}\label{KSgv}
   \alpha  \vv{x} = \mtr{R}(\qq{v})\,\vv{c},
\end{equation}
where $\mtr{R}$ is defined in eq. (\ref{Rotm}).
Alternatively, we can represent (\ref{KSgv}) as
\begin{equation}\label{KSgv:v}
    \alpha \vv{x} = \left( v_0^2 - \vv{v}\cdot \vv{v} \right) \, \vv{c} + 2 \left(\vv{c}\cdot \vv{v}\right) \vv{v}
    + 2 v_0 \vv{v} \times \vv{c},
\end{equation}
or
\begin{equation}\label{KSgv:h}
\alpha \vv{x} =  \left(\vv{c} \cdot \vv{v}\right)\,\vv{v}+\left[\qq{v} \wedge (\qq{v} \wedge \qq{c})\right]^\natural.
\end{equation}

The fibration property, known since \citet{Kust:64}, may be stated in the general case as follows:
quaternions $\qq{v}$ and
\begin{align}
\qq{w} & = \qq{v}\,(\cos{\phi}, \sin{\phi}\, \vv{c}) = \nonumber \\
 & = \left( v_0 \cos{\phi} - (\vv{v}\cdot \vv{c}) \sin{\phi}, \,
\cos{\phi}\, \vv{v} + ( v_0 \vv{c} + \vv{v} \times \vv{c} ) \sin{\phi} \right),
\label{fib}
\end{align}
generate the same vector $\vv{x}$ for all values of angle $\phi$. In other words, the point $\vv{x}$ maps onto
a fiber consisting of all quaternions $\qq{w}$ generated from a given representative $\qq{v}$.
The proof is elementary, once we recall equation (\ref{coru}). Then
\begin{equation}\label{prf:1}
    \qq{w}(0,\vv{c}) \bar{\qq{w}} = \qq{v} \left[ (\cos{\phi}, \sin{\phi}\, \vv{c})\,(0,\vv{c}) \,
 (\cos{\phi}, - \sin{\phi} \, \vv{c}) \right] \bar{\qq{v}} = \qq{v}  (0,\vv{c})   \bar{\qq{v}},
\end{equation}
because the part in square brackets describes the rotation of $\vv{c}$ around itself.
Thus, to a given rotation/scaling matrix $\mtr{R}(\qq{v})$, exactly two quaternions can be assigned ($\qq{v}$ and
$-\qq{v}$), but the product of $\mtr{R}$ and a specified vector $\vv{c}$ allows more freedom.
This means also that the `geometrical interpretation', stated in Section~\ref{KSER}, refers to
only one representative of the fiber.

According to the fibration property, inverting the transformation (\ref{KSgv}) amounts to
picking up some particular $\qq{v}$ that serves as the generator of the fiber. Since $\mtr{R}$ is
homogenous of degree 2, introducing a unit quaternion $\qq{q} = \qq{v}/|\qq{v}|$ we obtain
\begin{equation}\label{KSgv:1}
   \frac{\alpha  \vv{x}}{|\qq{v}|^2} = \mtr{R}(\qq{q})\,\vv{c},
\end{equation}
where $\mtr{R}(\qq{q}) \in \mathrm{SO}(3,\mathbb{R})$. Accordingly
\begin{equation}\label{rgen}
   || \vv{x} || = r = \frac{\qq{v} \cdot \qq{v}}{\alpha},
\end{equation}
since $||\vv{c}||=1$ by the assumptions. Thus, the transformation
\begin{equation}\label{KSgv:2}
    \hat{\vv{x}} =  \mtr{R}(\qq{q})\,\vv{c}, \qquad  \mbox{where}  \quad \hat{\vv{x}} = \frac{\vv{x}}{r},
\end{equation}
has the meaning of rotation from $\vv{c}$ to $\hat{\vv{x}}$.

Recalling the axis-angle decomposition (\ref{qrot}) we can aim at some `natural' choice of
$\qq{q}$ based upon rotation axis $\vv{n}$ and angle $\theta$ resulting from elementary
vector identities. Thus, the vector part of $\qq{q}$ is
\begin{equation}\label{cs:a}
   \vv{q} =  \sin{\frac{\theta}{2}} \, \vv{n} = \frac{\vv{c} \times \hat{\vv{x}}}{2\,q_0},
\end{equation}
whereas,
\begin{equation}\label{cs:b}
    q_0 = \cos{\frac{\theta}{2}} =   \sqrt{\frac{1+\vv{c} \cdot \hat{\vv{x}}}{2}}.
\end{equation}
Note the singular case $\vv{c} \cdot \hat{\vv{x}} = - 1$, when the actual choice should be
$\qq{q} = (0,\vv{n})$, with an arbitrary unit vector $\vv{n}$ orthogonal to $\vv{c}$ (and thus to $\hat{\vv{x}}$).
For a (perturbed) Kepler problem, it may only happen on a polar orbit, with $\vv{c}$ placed in the (osculating) orbit plane.
Another problematic situation concerns the collision $\vv{x}=\vv{0}$, when the KS quaternion $\qq{v}=\qq{0}$
simply cannot be normalized to $\vv{q}$, and the notion of rotation is inappropriate.

Thus we first propose an inversion rule
\begin{equation}\label{inv:1}
    \qq{v} = \sqrt{\frac{\alpha}{2}}\,\left( \sqrt{ r+\vv{c} \cdot \vv{x}} , \,
    \frac{\vv{c} \times \vv{x}}{ \sqrt{  r+\vv{c} \cdot \vv{x}} } \right), \qquad
    \mbox{for} \quad \vv{c} \cdot \hat{\vv{x}} > -1,
\end{equation}
and
\begin{equation}\label{inv:2}
    \qq{v} = \sqrt{\alpha r}\,\left( 0, \vv{n} \right), \quad \vv{n} \cdot \vv{x}=0, \qquad \mbox{for} \quad
    \vv{c} \cdot \hat{\vv{x}} = -1.
\end{equation}
It differs from the rules of \citet{StS:71}, effectively based upon the sign of $\vv{c} \cdot \vv{x}$ (with $\vv{c} = \vv{e}_1$).

An interesting alternative was proposed by \citet{Saha:09}, who used the inversion rule implying
a pure vector $(0,\vv{v})$ form. We can obtain it from  (\ref{inv:1})
through a quaternion product of $\qq{v}$ with $(0, \vv{c})$ or its conjugate (both being particular cases of (\ref{fib})).
The result is indeed considerably simpler:
\begin{equation}\label{inv:3}
    \qq{v} = \pm \sqrt{\frac{\alpha}{2 r \left(    1 +\vv{c} \cdot \hat{\vv{x}} \right)} } \,\left( 0 , \,
     \vv{x} + r \vv{c}  \right), \qquad
    \mbox{for} \quad \vv{c} \cdot \hat{\vv{x}} > -1,
\end{equation}
and (\ref{inv:2}) otherwise. The sign plus or minus can be chosen at will (both $\qq{v}$ and $-\qq{v}$ belong to the same fiber).
In practice, while converting
a sequence of positions forming an orbit $\vv{x}(t)$, we can swap the signs at the instants, where
the motion in $\qq{v}$ appears discontinuous. Note that the choice (\ref{inv:3}) allows plotting
the evolution of  KS variables in $\mathbb{R}^3$, furnishing spatial trajectory $\vv{v}(t)$.
The vector $\vv{v}$ defined by (\ref{inv:3}) will be called an SKS vector (after Saha, Kustaanheimo, and Stiefel).
In order to distinguish general formulae from the ones referring to the SKS vector, we will use the subscript
`s' for the latter.

Any KS quaternion $\qq{v}$ with $v_0 \neq 0$ can be reduced to the SKS vector by the product
\begin{equation}\label{redu}
    (0,\vv{v}_\mathrm{s}) = \pm \, \qq{v} \, \qq{q}_\mathrm{s},
\end{equation}
where the gauge function is a unit quaternion
\begin{equation}\label{qs}
    \qq{q}_\mathrm{s} = \frac{\left(\vv{v} \cdot  \vv{c}, \,  v_0 \vv{c} \right)}{\sqrt{v_0^2 + (\vv{v}\cdot \vv{c})^2}}.
\end{equation}
This reduction rule can be used if KS coordinates are followed without reference to the Cartesian position $\vv{x}$.

\subsection{Canonical extension}

\subsubsection{KS momenta}

Being interested in Hamiltonian formulation of the Kepler problem, we need to match the KS variables $\qq{v}$
with their conjugate momenta $\qq{V}$. This goal can be achieved by a dimension raising Mathieu transformation
$\vv{X}\cdot\rd\vv{x} = \qq{V} \cdot \rd \qq{v}$, where $\vv{X}$ are the momenta conjugate to the Cartesian $\vv{x}$
coordinates. We will also use a formal quaternion $\qq{X}$, postulating $X_0 = 0$. Following the standard procedure
\citep{Kur:77,DEF}, we generalize it from $\qq{e}_1$ to a unit quaternion $\qq{c} = (0,\vv{c})$ obtaining
\begin{equation}\label{VtoXq}
    \qq{X} = \frac{\qq{V} \qq{c} \bar{\qq{v}}}{2r},
\end{equation}
wherefrom the vector part is
\begin{align}
\vv{X} & = \frac{(\vv{c}\cdot\vv{v})\,\vv{V} + \left[\,\qq{V} \wedge (\qq{v} \wedge \qq{c} ) \right]^\natural}{2r} =
\nonumber \\
 & = \frac{1}{2r} \left[ (\vv{c} \cdot \vv{v})\, \vv{V}+(v_0 V_0 - \vv{v} \cdot \vv{V})\, \vv{c} + ( V_0 \vv{v} +
   v_0 \vv{V} ) \times \vv{c} + (\vv{c} \cdot \vv{V})\, \vv{v} \right]. \label{VtoX}
\end{align}
Since the scalar component of (\ref{VtoXq}) should be null, we obtain the constraint
\begin{equation}\label{invarc}
   X_0 =   \frac{\vv{J} \cdot \vv{c}}{2r} = 0, \qquad \mbox{where} \qquad
    \vv{J} =  -v_0 \vv{V}+ V_0 \vv{v} +  \vv{v} \times \vv{V}.
\end{equation}

Remarkably the vector $\vv{J}$, orthogonal to $\vv{c}$, is directly related to the quaternion cross product (\ref{qop})
\begin{equation}\label{Jop}
  \qq{J} = (0,\vv{J}) =  \bar{\qq{v}}  \wedge  \bar{\qq{V}},
\end{equation}
so  we can rewrite the condition (\ref{invarc}) as
\begin{equation}\label{invarq}
    X_0 = \frac{\left(  \bar{\qq{v}}  \wedge  \bar{\qq{V}}  \right) \cdot \qq{c}}{2 r} = 0,
\end{equation}
valid regardless of $c_0$. This is the general equivalent of the KS1 phase space constraint
(\ref{ivmS}) or (\ref{inv2}) for the transformation (\ref{KSgen}).
Using $\qq{c}=\qq{e}_1$, we can recover the formula
of \citet{DEF},
\begin{equation}\label{Jdef}
 \vv{J} \cdot \vv{e}_1 =   \qq{v} \cdot (\qq{V} \qq{e}_1) = 0.
\end{equation}

The presence of constraint (\ref{invarq}) allows a unique determination
of $V_0$ in terms of the remaining variables. This, combined with a possibility
of reduction to $\vv{v}_\mathrm{s}$ and $\qq{V}_\mathrm{s}$ implies that in spite
of using eight variables, we follow the dynamics of a system with
effectively three degrees of freedom.

The inverse of the transformation (\ref{XtoVq}) is given as a quaternion product
\begin{equation}\label{XtoVq}
    \qq{V} = \frac{2\,\qq{X} \,\qq{v} \,\bar{\qq{c}}}{\alpha},
\end{equation}
or, explicitly (setting $X_0=0$)
\begin{align}
     V_0 & = \frac{2\,\left( \bar{\qq{v}} \wedge \qq{X} \right) \cdot \qq{c}}{\alpha} =
     \frac{2 \, (v_0 \vv{c} + \vv{v} \times \vv{c} )\cdot \vv{X}}{\alpha}, \nonumber \\
  \vv{V} & = \frac{2}{\alpha} \left[
    (\vv{c} \cdot \vv{v})\,\vv{X} + (\vv{v} \cdot \vv{X}) \, \vv{c}  - (\vv{c} \cdot \vv{X}) \, \vv{v}   +  v_0 (\vv{c} \times \vv{X})\right].
    \label{Xtov}
\end{align}

Notably, the definition of the new momenta is given in a mixed form, involving old momenta $\vv{X}$ and
new coordinates $\vv{v}$. It means, that for a given set of values $\vv{x},\vv{X}$, the phase space fiber
contains not only the family of coordinates $\qq{v}$ implied by (\ref{fib}), but also the family of momenta
$\qq{V}$ -- one quaternion for each member of (\ref{fib}). Thus, if we try to provide the explicit form
of $\qq{V}(\vv{x},\vv{X})$, we may choose some particular representative of the fiber. Let us comfortably
choose the SKS vector (\ref{inv:3}), because then
we can simplify expressions, remaining on the ground of usual vector calculus.

In the absence of $v_0$, the scalar part $V_0$ simplifies to
$V_{0\mathrm{s}} = - 2 (\vv{v}_\mathrm{s} \times \vv{X}) \cdot \vv{c}/\alpha$.
Substituting $\vv{v}_\mathrm{s}$ from (\ref{inv:3}), with the plus sign selected, into the first of equations (\ref{Xtov}),
we find
\begin{equation}\label{intV0}
   V_{0\mathrm{s}} =    - \sqrt{\frac{2}{\alpha \, r \left(    1+\vv{c} \cdot \hat{\vv{x}} \right)} } \,( \vv{x} \times \vv{X} ) \cdot \vv{c},
\end{equation}
whereas for $\hat{\vv{x}}= -\vv{c}$, the scalar part of $\qq{V}$ is $V_{0\mathrm{s}}=0$. In both cases the conclusion is
the same: whenever the SKS vector is taken for coordinates,
the scalar part of the KS momenta quaternion is a product of a coordinates dependent factor and the
projection of angular momentum on the defining vector $\vv{c}$.

The vector part $\vv{V}$ is also linked with familiar quantities when the same SKS vector is used, leading to
\begin{equation}\label{intV}
   \vv{V}_{\mathrm{s}} =     \sqrt{\frac{2}{\alpha\, r \left(    1+\vv{c} \cdot \hat{\vv{x}} \right)} }
   \,\left( r \vv{X} + (\vv{x}\cdot \vv{X})\,\vv{c} +  (\vv{x} \times \vv{X} ) \times \vv{c} \right),
\end{equation}
where the Cartesian momentum, radial velocity and angular momentum appear.

Finding $\qq{V}_\mathrm{s}$ is possible without the knowledge of $\vv{x}$ and $\vv{X}$. Let us multiply
both sides of eq.~(\ref{XtoVq}) by a quaternion product $\qq{c} \qq{q}_\mathrm{s} \bar{\qq{c}}$.
Then, the equation becomes
\begin{equation}\label{redV}
 \qq{V} \qq{c} \qq{q}_\mathrm{s} \bar{\qq{c}} = \frac{2\,\qq{X} \,(\qq{v}\,\qq{q}_\mathrm{s} )\,\bar{\qq{c}}}{\alpha},
\end{equation}
and we see that its left hand side should represent $\qq{V}_\mathrm{s}$. But one may easily verify that
$\qq{c} \qq{q}_\mathrm{s} \bar{\qq{c}} = \qq{q}_\mathrm{s}$, so if momenta $\qq{V}$ are determined by
$\vv{X}$ with an arbitrary quaternion $\qq{v}$, then the momenta $\qq{V}_\mathrm{s}$ determined by the same
$\vv{X}$ and the equivalent SKS vector $\vv{v}_\mathrm{s}$ are
\begin{equation}\label{redV1}
    \qq{V}_\mathrm{s} = \pm \qq{V}\, \qq{q}_\mathrm{s},
\end{equation}
where the sign choice should be the same as in (\ref{redu}).

\section{KS Hamiltonian and its invariants}
\label{hamil}

\subsection{Perturbed Hamiltonian and equations of motion}
\label{mam}

As long as the defining vector is constant, the canonical KS transformation
is time independent, so the Hamiltonian function transforms without a remainder.
Thus, the first step is to obtain $\cH^\star(\qq{v},\qq{V},t) = \cH(\vv{x},\vv{X},t)$,
where $\cH$ is a perturbed two body problem Hamiltonian
\begin{equation}\label{Hgen}
    \cH(\vv{x},\vv{X},t) = \cH_0(\vv{x},\vv{X})  + \mathcal{R}(\vv{x},\vv{X},t),
\end{equation}
with  the Keplerian part
\begin{equation}\label{Hkep}
    \cH_0 = \frac{\vv{X} \cdot \vv{X}}{2} - \frac{\mu}{r},
\end{equation}
depending on the gravitational parameter $\mu$, and we make no assumptions about the
order of magnitude for the perturbation $\mathcal{R}$.

The distance $r$ may be considered a known function of $\vv{v}$ thanks to (\ref{rgen}),
so we retain this symbol in $\cH_0$.
The square of $|\vv{X}|$ is easily found from $\qq{X}\bar{\qq{X}} = \vv{X} \cdot \vv{X} + X_0^2$. Substituting
eq. (\ref{VtoXq}), using the rule (\ref{coru}) and defining $X_0$ by (\ref{invarc}), we find\footnote{The equivalent
equation (22) of \citet{DEF} is incomplete by the omission of $X_0^2$.}
\begin{equation}\label{X2}
 \vv{X} \cdot \vv{X} = \frac{\alpha}{4 r} \qq{V} \cdot \qq{V} -\frac{(\vv{J} \cdot \vv{c})^2}{4 r^2}.
\end{equation}
Thus the transformed Keplerian Hamiltonian is
\begin{equation}\label{Hkep:1}
 \cH^\star_0(\qq{v},\qq{V}) = \frac{\alpha}{8 r} \qq{V} \cdot \qq{V} - \frac{\mu}{r} - \frac{(\vv{J} \cdot \vv{c})^2}{8 r^2}.
\end{equation}

Dropping the last, zero valued term in (\ref{Hkep:1}) is allowed, but it should not be done without reflection.
It is to be remembered that while $x_0=0$ is the property of the point transformation (\ref{KSgen}) itself,
$X_0=0$ is only postulated. Two separate questions should be addressed. First: is the value
of $\vv{J}\cdot \vv{c}$ conserved during the motion? Here the answer is conditionally positive: it is not so for
an arbitrarily invented Hamiltonian function of $\qq{v}$ and $\qq{V}$. But if the Hamiltonian is a transformed
function $\cH(\vv{x},\vv{X},t)$, then its KS image $\cH^\star(\qq{v},\qq{V},t)$ conserves $\vv{J}\cdot \vv{c}$, because all
the Poisson's brackets $\pb{\vv{J}\cdot \vv{c}}{x_j}=\pb{\vv{J}\cdot \vv{c}}{X_j}=0$, for $0 \leq j \leq 3$,
similarly to the argument of \citet{DEF}.
The second question, less often considered, is: does the presence of $\vv{J} \cdot \vv{c}$ influence the
solution? Worth asking, because a zero valued function may still have nonzero derivatives. For a while the answer is
obvious, because $\cH_0^\star$ contains the square   $(\vv{J} \cdot \vv{c})^2$, so its gradient will vanish. But the problem
may reappear in the context of variational equations or when the rotating reference frame will be considered.

The Hamiltonian (\ref{Hkep}) remains singular at $r=0$, so we proceed with the Sundman transformation
(\ref{sund-K})
\begin{equation}\label{Sund}
    \frac{\rd \tau}{\rd t} = \frac{\beta}{r},
\end{equation}
with an arbitrary parameter $\beta$. In the canonical framework, switching from physical time $t$ to the Sundman time $\tau$
requires the transition to the extended phase space, appending to the KS coordinates and momenta a conjugate pair
$v^\ast$ and $V^\ast$. The former is an imitator of the physical time (up to an additive constant); the latter
serves to fix a zero energy manifold for the motion, being a doppelganger of the Hamiltonian $\cH^\ast$.
Thus, on the manifold $\cH^\star+V^\ast =0$, we can divide the extended Hamiltonian by the right hand side of
(\ref{Sund}), obtaining the Hamiltonian function
\begin{equation}\label{cK}
    \cK(\qq{v},\qq{V},v^\ast,V^\ast) = \frac{\alpha}{8 \beta} \qq{V} \cdot \qq{V} - \frac{\mu}{\beta} +
     \frac{r}{\beta} \mathcal{R}^\star(\qq{v},\qq{V},v^\ast) + \frac{V^\ast r}{\beta} = 0,
\end{equation}
independent on the new time variable $\tau$. What remains, is a judicious choice of $\alpha$ and $\beta$.
Assuming
\begin{equation}\label{beta}
    \beta = \frac{\alpha}{4}, \qquad \mbox{hence} \quad \frac{\rd \tau}{\rd t} = \frac{\alpha}{4 r},
\end{equation}
we secure $\qq{v}' = \qq{V}$. If then $\alpha$ is equal to major axis (i.e. $\alpha = 2 a$) of the elliptic orbit, $\tau$
will run on average at half rate of $t$ for the Kepler problem, in accord with the angle doubling
property of the KS transformation.

Although the constant term has no influence on equations of motion, we retain it for its role in fixing the $\cK=0$
manifold. And so we finally set up the KS Hamiltonian
\begin{align}\label{Kham}
\cK(\qq{v},\qq{V},v^\ast,V^\ast) & = \cK_0(\qq{v},\qq{V},V^\ast)+ \mathcal{P}(\qq{v},\qq{V},v^\ast), \\
\cK_0(\qq{v},\qq{V},V^\ast) & = \frac{1}{2} \,\qq{V} \cdot \qq{V} +  \frac{4 V^\ast }{\alpha^2} \qq{v} \cdot \qq{v} - \frac{4\mu}{\alpha}, \\
\mathcal{P}(\qq{v},\qq{V},v^\ast) & = \frac{4 r}{\alpha} \mathcal{R}^\star(\qq{v},\qq{V},v^\ast).
\end{align}
The value of $V^\ast$ should secure $\cK = 0$.

Equations of motion resulting from (\ref{Kham}) are those of a perturbed harmonic oscillator with  frequency
\begin{equation}
\omega_0 =  \frac{2 \sqrt{2 V^\ast}}{ \alpha},
\end{equation}
namely
\begin{align}
\qq{v}' & = \pb{\qq{v}}{\cK}  = \qq{V} + \sum_{j=0}^3 \frac{\partial \mathcal{P}}{\partial V_j} \qq{e}_j, \\
\qq{V}' & = \pb{\qq{V}}{\cK}  = - \omega_0^2\,\qq{v} - \sum_{j=0}^3 \frac{\partial \mathcal{P}}{\partial v_j} \qq{e}_j, \\
(v^\ast)' & = \pb{v^\ast}{\cK}  = \frac{4r}{\alpha}, \\
(V^\ast)' & = \pb{V^\ast}{\cK}  = \frac{\partial \mathcal{P}}{\partial v^\ast},
\end{align}
where the prime marks the derivative with respect to Sundman time $\tau$.
We can note, that using an arbitrary defining vector $\vv{c}$ has no influence on the unperturbed problem in KS variables.
Whatever change results from a particular choice of $\vv{c}$, may be revealed only by the form taken
by $\mathcal{P}$ in a specific problem.

\subsection{Invariants}
\label{invar}

Including the perturbation $\mathcal{P}$, we can only mention two invariants: if $\mathcal{P}$ does not depend
explicitly on  $v^\ast$ (hence on time $t$), the momentum $V^\ast = - E = \mathrm{const}$, where $E$ is the
total energy; regardless of $\mathcal{P}$, the scalar product $\vv{J}\cdot\vv{c}=0$ is also invariant, as already mentioned.
Thus, let us focus on the first integrals of the unperturbed system $\cK_0$, i.e. the two body problem.

There are two points of view for the unperturbed system. We can see it as a four dimensional isotropic harmonic oscillator
with its own first integrals. But we can also see it as a transformed Kepler problem with the well known
first integrals, potentially expressible in terms of the oscillator constants.

\subsubsection{Oscillator}

In the absence of perturbation, we can consider the Hamiltonian $\cK_0$  as a separable system of four independent oscillators
\begin{equation}\label{NHp}
\cK_0 = - \frac{4\mu}{\alpha}+ \sum_{j=0}^3 \mathcal{N}_j=0, \qquad \mbox{where} \quad \mathcal{N}_j = \frac{V_j^2}{2}+ \frac{\omega_0^2 v_j^2}{2}=E_j.
\end{equation}
Each Hamiltonian $\mathcal{N}_j$ is a first integral in involution with the rest $\pb{\mathcal{N}_i}{\mathcal{N}_j}=0$,
but only three of them are independent, since their sum is fixed by (\ref{NHp}).
Yet, the system is superintegrable and more integrals can be found. First, we can introduce a four-dimensional variant of the Fradkin tensor
$\mtr{F}$ \citep{Fra:67}
\begin{equation}\label{FT}
    F_{ij} = \frac{V_i V_j}{\omega_0} + \omega_0 v_i v_j.
\end{equation}
The symmetric matrix $\mtr{F}$ contains 10 different first integrals (not all independent),
including four diagonal terms $F_{ii} = 2E_i/\omega_0$.

Another set of integrals constitutes an antisymmetric angular momentum matrix $\mtr{L}$
\begin{equation}\label{AMM}
    L_{ij} = v_i V_j - v_j V_i,
\end{equation}
with 6 distinct elements (again, not all independent).

Considering $F_{ij}$ and $L_{ij}$ as the generators of Hamiltonian equations, we signal their different, somewhat complementary roles.
Equations
\begin{equation}\label{gen:F}
    \qq{v}'  = \pb{\qq{v}}{F_{ij}} = \frac{V_j \qq{e}_i + V_i \qq{e}_j}{\omega_0}
    , \qquad \qq{V}'  = \pb{\qq{V}}{F_{ij}} = - \omega_0 \left( v_j \qq{e}_i + v_i \qq{e}_j \right),
\end{equation}
define a phase plane rotation: either in a phase plane $(v_i,V_i/\omega_0)$ (diagonal terms $F_{ii}$), or
in two phase planes $(v_i,V_i/\omega_0)$ and $(v_j,V_j/\omega_0)$. The angular momentum terms $L_{ij}$
lead to equations
\begin{equation}\label{gen:L}
    \qq{v}'  = \pb{\qq{v}}{L_{ij}}= - v_j \qq{e}_i + v_i \qq{e}_j, \qquad \qq{V}'  = \pb{\qq{V}}{L_{ij}}
    =  - V_j \qq{e}_i + V_i \qq{e}_j,
\end{equation}
that generate rotation on a coordinate plane $(v_i,v_j)$ and on a momentum plane $(V_i,V_j)$.

Two important cross products that appeared in the KS transformation are expressible in terms of
$L_{ij}$ and thus are first integrals of unperturbed motion:
\begin{align}
\left( \qq{v} \wedge \qq{V} \right)^\natural & = \left( L_{01} + L_{23}\right) \vv{e}_1 +  \left( L_{02} + L_{31}\right) \vv{e}_2
+ \left( L_{03} + L_{12}\right) \vv{e}_3, \label{auxL} \\
\left( \bar{\qq{v}} \wedge \bar{\qq{V}} \right)^\natural & =  \left( L_{10} + L_{23}\right) \vv{e}_1 + \left( L_{20} + L_{31}\right) \vv{e}_2
 + \left( L_{30} + L_{12}\right) \vv{e}_3.
\end{align}
Accordingly, the condition (\ref{invarq}) can be seen as a constraint on the angular momentum of the oscillator.

\subsubsection{Kepler problem}
\label{kepro}

The energy integral of the Kepler problem has been already discussed in Section~\ref{hamil},
so let us pass to the two vector-valued first integrals: angular momentum and Laplace (Runge-Lenz) vector.

Expressing the angular momentum
\begin{equation}\label{angmom}
    \vv{G} = \vv{x} \times \vv{X},
\end{equation}
in terms of the KS variables is a formidable task if a brute force attack is attempted by substitution of
(\ref{KSgv:v}) and the second line of (\ref{VtoX}) into (\ref{angmom}). It is much better to start from
plugging in a cross product $\qq{x} \wedge \qq{X}$, so that
\begin{equation}\label{am:01}
\vv{x} \times \vv{X} = \left[\,\qq{x} \wedge \qq{X}\right]^\natural + X_0 \vv{x},
\end{equation}
with $x_0=0$.
Substituting the quaternion product forms (\ref{KSgen}) for $\qq{x}$ and (\ref{VtoXq}) for $\qq{X}$, one can resort to the
factor exchange rule (\ref{cper})
\begin{equation}
    \qq{x} \wedge \qq{X}  = \frac{\left(\qq{v} \qq{c} \bar{\qq{v}}\right) \wedge \left(\qq{V} \qq{c} \bar{\qq{v}}\right)}{2 \alpha r}
    = \frac{\left(\qq{v} \qq{c}  \right) \wedge \left(\qq{V} \qq{c} \bar{\qq{v}} \qq{v} \right)}{2 \alpha r}
    = \frac{ \qq{v}  \wedge \left(\qq{V} \qq{c} \bar{\qq{c}} \right)}{2} = \frac{ \qq{v}   \wedge  \qq{V} }{2},
\label{am:02}
\end{equation}
where we used $\bar{\qq{v}}\qq{v}= \alpha r$, and $\qq{c} \bar{\qq{c}} = 1$.
Thus we obtain
\begin{equation}\label{am:fin}
 \vv{G} = \vv{x} \times \vv{X} = \frac{\left[\,\qq{v} \wedge \qq{V}\right]^\natural}{2}+ X_0 \vv{x}
 = \frac{1}{2} \left[  \,\qq{v} \wedge \qq{V}   + \frac{\vv{J} \cdot \vv{c}}{\alpha r}\, \qq{v} \qq{c} \bar{\qq{v}}
 \right]^\natural .
\end{equation}
An analogous expression was obtained by \citet{DEF} with a statement:
`Proof. - By straightforward calculation using Symbol Processor'.  We have decided to provide the proof
in full length (or rather shortness), as a good example of the situation where quaternion formalism beats standard vector calculus.

If the expression (\ref{am:fin}) is used for the sole purpose of computing the value of
$\vv{G}$, then $X_0=0$ can be safely set. But if $\vv{G}$ is to appear in the perturbing Hamiltonian
$\mathcal{P}$, one should not forget that derivatives of $X_0$ do not vanish in general.

Linking the Keplerian $\vv{G}$ with the oscillator's angular momentum tensor $\mtr{L}$ is straightforward: inserting
(\ref{auxL}) into (\ref{am:fin}) results in
\begin{equation}\label{am:L}
 \vv{G}
 = \frac{ L_{01} + L_{23}}{2} \vv{e}_1 +  \frac{ L_{02} + L_{31}}{2} \vv{e}_2
+ \frac{ L_{03} + L_{12}}{2} \vv{e}_3 + X_0 \vv{x}.
\end{equation}
Notably, the defining vector $\vv{c}$ has no direct effect on the direction of the angular momentum. Its action is only indirect,
through the constraint $\vv{J}\cdot \vv{c}=0$.

The Laplace vector $\vv{e}$ is primarily given with  a cross product of momentum and angular momentum, but in order to express
it in terms of KS variables, a version resulting form the `BAC-CAB' identity is more convenient
\begin{equation}\label{Lv:def}
    \mu \vv{e} = \vv{X} \times \vv{G} - \mu \hat{\vv{x}} = \left( \vv{X} \cdot \vv{X} - \mu r^{-1}\right)\, \vv{x} - (\vv{x} \cdot \vv{X})\,\vv{X}.
\end{equation}
Most of the building blocks are ready in equations (\ref{KSgv:v}), (\ref{KSgv:h}), (\ref{VtoX}), and (\ref{X2}). The remaining
product is elementarily found applying the identity (\ref{dper})
\begin{equation}\label{xX}
    \vv{x} \cdot \vv{X} = \qq{x}\cdot \qq{X} = \frac{(\qq{v} \qq{c} \bar{\qq{v}}) \cdot  (\qq{V} \qq{c} \bar{\qq{v}})}{2 \alpha r}
    = \frac{\bar{\qq{V}} \cdot \left(\qq{c} \bar{\qq{v}}  \overline{(\qq{v} \qq{c} \bar{\qq{v}})} \right) }{2 \alpha r}
    = \frac{\bar{\qq{V}} \cdot \left(\qq{c} \bar{\qq{v}}   \qq{v} \bar{\qq{c}}  \bar{\qq{v}}  \right) }{2 \alpha r} =
   \frac{\bar{\qq{V}} \cdot \bar{ \qq{v}} }{2  }  = \frac{\qq{v} \cdot  \qq{V}  }{2 }.
\end{equation}
And so the Laplace vector can be expressed as a quaternion product
\begin{equation}\label{Lap:KS}
    \mu \vv{e} = \frac{1}{2r}\left[ \left( \left( \frac{\qq{V} \cdot \qq{V}}{2}- \frac{2\mu}{\alpha}
    + \frac{(\vv{J} \cdot \vv{c})^2}{2 \alpha r}\right) \qq{v}
    - \frac{\qq{v} \cdot \qq{V}}{2} \qq{V} \right) \qq{c} \bar{\qq{v}} \right]^\natural.
\end{equation}
The formula does not look friendly, but one should expect it to be expressible in terms of Fradkin integrals.
Indeed, after rather tedious manipulations, we have found that
\begin{equation}\label{Lap:FT}
    \mu \vv{e} =  - \frac{\alpha \omega_0}{4}\,\mtr{E} \vv{c} - X_0 \vv{G} + \frac{\alpha \cK_0}{4 r} \vv{x},
\end{equation}
where the second term to the right vanishes due to the KS constraint $\vv{J}\cdot \vv{c}=0$, and the third is null on the Keplerian manifold
$\cK_0=0$. It is not by chance, that the form of matrix $\mtr{E}$
\begin{equation}\label{Em}
   \mtr{E}  = \left(
               \begin{array}{ccc}
                 E_{11} &  \quad   F_{12} - F_{03}   &  \quad  F_{13} + F_{02}   \\
                    F_{12} +  F_{03}  & E_{22} &  \quad F_{23} - F_{01}    \\
                  F_{13} -  F_{02}    &  \quad   F_{23} +  F_{01}   & E_{33} \\
               \end{array}
             \right),
\end{equation}
with diagonal terms
\begin{eqnarray}
  E_{11} &=& \frac{ F_{00} + F_{11} - F_{22} - F_{33}}{2} = \frac{\cK_0}{\omega_0} + \frac{4\mu}{\alpha \omega_0} -  F_{22}- F_{33},   \nonumber \\
  E_{22} &=&  \frac{F_{00} - F_{11} + F_{22} - F_{33}}{2} = \frac{\cK_0}{\omega_0} + \frac{4\mu}{\alpha \omega_0} -  F_{11}- F_{33},
  \label{Ediag} \\
  E_{33} &=& \frac{F_{00} - F_{11} - F_{22} + F_{33}}{2} = \frac{\cK_0}{\omega_0} + \frac{4\mu}{\alpha \omega_0} -  F_{11} - F_{22}, \nonumber
\end{eqnarray}
mimics the matrix $\mtr{R}(\vv{v})$ defined by eq.~(\ref{Rotm}), present in the KS transformation formula (\ref{KSgv}).
In contrast to the angular momentum $\vv{G}$, the defining vector $\vv{c}$ is explicitly present in the definition of $\vv{e}$.

\subsection{Dynamical role of the invariant $\vv{J}\cdot \vv{c}$}
\label{Jcdyn}

In Sections~\ref{mam} and \ref{kepro}, some warnings have been issued concerning the presence of the invariant $\vv{J}\cdot \vv{c}=0$,
which should not be dropped blindly in some expressions. Let us now inspect its influence, by considering a Hamiltonian
$\mathcal{M} = \Psi \vv{J}\cdot \vv{c}$, where $\Psi$ is an arbitrary function of KS variables.
Canonical equations of motion generated by $\mathcal{M}$, can be cast into a quaternion product form
\begin{equation} \label{genM}
  \qq{v}' = \left\{ \qq{v}, \mathcal{M} \right\} = \Psi \,\qq{v} \bar{\qq{c}}, \qquad
  \qq{V}'  = \left\{ \qq{V}, \mathcal{M} \right\} =  \Psi \,\qq{V} \bar{\qq{c}}.
\end{equation}
The solution of this system is a quaternion product
\begin{equation}\label{solM}
    \qq{v} = \qq{u} \left( \cos{\phi}, \sin{\phi} \,\vv{c} \right), \qquad \qq{V} = \qq{U} \left( \cos{\phi}, \sin{\phi}\, \vv{c} \right),
\end{equation}
with arbitrary constants $\qq{u}, \qq{U}$, provided $\phi$ is a function of time (possibly implicit) satisfying $\phi'= - \Psi$.
But, recalling the fiber definition (\ref{fib}), we see that the resulting evolution of KS variables (\ref{solM})
happens on a fiber referring to constant values of the Cartesian variables.

Considering any Hamiltonian $\mathcal{K}+\mathcal{M}$, where $\cK$ is a KS transform of some function of the Cartesian variables,
we recall that both terms commute ($\left\{\cK,\mathcal{M} \right\} = 0$), so the the solution will be a direct composition of
flows resulting from both Hamiltonians separately. As it follows, adding or retracting a $\vv{J}\cdot \vv{c}$ term to a Hamiltonian function
can modify a trajectory in the KS phase space,
but it has no influence on the resulting motion in the Cartesian variables $\vv{x},\vv{X}$.
We encourage an interested reader to consult a related work of \citet{RUP:16} in the framework of the KS1 set.

\section{Kepler problem in rotating reference frame}

\label{keprot}
\subsection{General equations of motion}

The Kepler problem in a uniformly rotating reference frame is a necessary building block for a number of dynamical problems
handled by analytical perturbation techniques or symplectic integrators with partitioned Hamiltonian. Recently, the problem has been
solved by \citet{LB:15}, where the account of earlier works can be found as well. Since \citet{LB:15} solved the problem in the frame rotating around $\vv{e}_3$
using the KS1 set (based upon the defining vector $\vv{c}=\vv{e}_1$), it may be interesting to confront the solution with a new one,
assuming an arbitrary direction of rotation axis and benefiting from the freedom in the defining vector choice.
Intuitively, selecting $\vv{c}$ directed along the rotation axis seems most appropriate, so  we assume
the angular velocity vector of the reference frame to be $\Omega \vv{c}$ from the onset.

The KS transformation, as described in Section~\ref{KSarb}, will be applied to the coordinates $\vv{x}$ and momenta $\vv{X}$
of the rotating frame. If at the epoch $t=0$ the rotating frame and the fixed frame axes coincide, then the transformation linking
$\vv{x},\vv{X}$ with the fixed frame coordinates $\vv{x}_\mathrm{f}$ and momenta $\vv{X}_\mathrm{f}$
\begin{equation}\label{rf}
    \vv{x} = \mtr{R}(\qq{q}) \vv{x}_\mathrm{f}, \qquad \vv{X} = \mtr{R}(\qq{q}) \vv{X}_\mathrm{f},
\end{equation}
involves rotation matrix from eq.~(\ref{Rotm}) with the rotation quaternion
\begin{equation}\label{rotq}
    \qq{q} = \left( \cos{\left(\frac{\Omega t}{2}\right)}, - \sin{\left(\frac{\Omega t}{2}\right)}\,\vv{c} \right).
\end{equation}
The transformation is canonical and, being time-dependent, it creates the remainder $-\Omega \vv{G} \cdot \vv{c}$ supplementing the
transformed Hamiltonian. Accordingly, the Hamiltonian function to be considered is a sum of $\cK_0$ from eq.~(\ref{Kham})
and of
\begin{equation}
\mathcal{P} = - \frac{4 r \Omega}{ \alpha} \, \vv{G} \cdot \vv{c}.
\end{equation}

\subsection{Simplification}

Recalling the conclusion of Section~\ref{Jcdyn}, and benefiting from the choice of $\vv{c}$,
we can modify $\mathcal{P}$ and use
\begin{equation}\label{Pp}
    \mathcal{P}_\mathrm{m} = \mathcal{P} - \frac{2  \Omega}{\alpha}
    \left(  \vv{x} \cdot \vv{c} + r \right)\,\vv{J} \cdot \vv{c}.
\end{equation}
According to equations (\ref{am:fin}), (\ref{auxL}) and (\ref{Jop}), the modified term is simply
\begin{equation}\label{Pm}
    \mathcal{P}_\mathrm{m} =   - \frac{4 r  }{\alpha} \Omega H, \qquad H = \left(\vv{v} \times \vv{V} \right) \cdot \vv{c},
\end{equation}
so it contains only the vector parts of the KS quaternions, save for $v_0$ present in $r$.

Deriving canonical equations of motion from
\begin{equation}\label{Krot}
    \cK = \frac{\qq{V} \cdot \qq{V}}{2}  +  \frac{4 V^\ast \, \qq{v} \cdot \qq{v} }{\alpha^2}  - \frac{4\mu}{\alpha}
     - \frac{4 \,\Omega \, \qq{v} \cdot \qq{v}  }{\alpha^2} \, \left(\vv{v} \times \vv{V} \right) \cdot \vv{c},
\end{equation}
we observe that they neatly split into a scalar part
\begin{equation}\label{eqm:s}
    v_0' = V_0, \qquad V_0' = - w^2 v_0,
\end{equation}
and the vector part
\begin{equation}\label{eqm:v}
    \vv{v}' = \vv{V} - \frac{4 r}{\alpha} \left(\Omega \vv{c} \times \vv{v} \right), \qquad \vv{V}' = - w^2 \vv{v}
      - \frac{4 r}{\alpha} \left(\Omega \vv{c} \times \vv{V} \right),
\end{equation}
where the frequency
\begin{equation}\label{wdef}
    w = \frac{2 \sqrt{2 (V^\ast - \Omega H)}}{\alpha},
\end{equation}
is a constant of motion, because $H' = \left\{H,\cK\right\}=0$, and the last two equations of motion are
\begin{equation}\label{eqm:ast}
    (v^\ast)' = \frac{4 r}{\alpha}, \qquad (V^\ast)' = 0.
\end{equation}
Initial conditions for this system at $\tau=0$ will be
\begin{equation}\label{inic}
    \qq{v}(0) = \qq{u}, \qquad \qq{V}(0) = \qq{U}, \qquad v^\ast(0) = t = 0.
\end{equation}

Thus, the situation is much more comfortable than in \citet{LB:15} and all earlier works. Equations~(\ref{eqm:s}),
describing a simple, one-dimensional harmonic oscillator, are easily
solved, rendering
\begin{equation}\label{sol:s}
    v_0 = \cos{(w \tau)} \,u_0 + \sin{(w \tau)} \, \frac{U_0}{w}, \qquad
    V_0 = - w \sin{(w \tau)} \, u_0 + \cos{(w \tau)} \, U_0,
\end{equation}
similarly to the fixed frame case.

Looking at the equations~(\ref{eqm:v}), we recognize two parts referring to harmonic oscillator dynamics
and to the kinematics of rotation. Introducing the cross product matrix
\begin{equation}\label{Cmat}
    \mtr{C} = \Omega \,\left(
                \begin{array}{ccc}
                  0 & -c_3 & c_2 \\
                  c_3 & 0 & -c_1 \\
                  -c_2 & c_1 & 0 \\
                \end{array}
              \right),
\end{equation}
we rewrite (\ref{eqm:v}) in the vector-matrix form
\begin{equation}\label{eqm:v1}
    \frac{\rd \vv{v}}{\rd \tau} = \vv{V} - \frac{4 r }{\alpha} \mtr{C} \vv{v},
    \qquad \frac{\rd \vv{V}}{\rd \tau}  = - w^2 \vv{v}
      - \frac{4 r  }{\alpha}\mtr{C} \vv{V},
\end{equation}
which suggest to postulate the solution
\begin{equation}\label{post:1}
    \vv{v} = \mtr{A}\, \left( b_1 \vv{u} + b_2 \vv{U}  \right), \qquad
    \vv{V} = \mtr{A}\, \left( b_3 \vv{u} + b_4 \vv{U}  \right),
\end{equation}
involving a common matrix $\mtr{A}$ and four scalars $b_j$, with the initial conditions
$\mtr{A}=\mtr{I}$, $b_2=b_3=0$, and $b_1=b_4=1$ at $\tau=t=v^\ast=0$. Substitution into (\ref{eqm:v1}) leads to
\begin{eqnarray}
    \frac{\rd \mtr{A}}{\rd \tau} \left( b_1 \vv{u} + b_2 \vv{U}  \right)  + \mtr{A}
    \left( \frac{\rd b_1}{\rd \tau} \vv{u} +  \frac{\rd b_2}{\rd \tau} \vv{U}\right)
    & = & \mtr{A}\, \left( b_3 \vv{u} + b_4 \vv{U}  \right) \nonumber \\
    & & - \frac{4 r  }{\alpha} \mtr{C}\mtr{A}\, \left( b_1 \vv{u} + b_2 \vv{U}  \right), \label{row:1}\\
        \frac{\rd \mtr{A}}{\rd \tau} \left( b_3 \vv{u} + b_4 \vv{U}  \right)  + \mtr{A}
    \left( \frac{\rd b_3}{\rd \tau} \vv{u} +  \frac{\rd b_4}{\rd \tau} \vv{U}\right)
    & = & - w^2 \mtr{A}\, \left( b_1 \vv{u} + b_2 \vv{U}  \right) \nonumber \\
    & & - \frac{4 r  }{\alpha} \mtr{C}\mtr{A}\, \left( b_3 \vv{u} + b_4 \vv{U}  \right). \label{row:2}
\end{eqnarray}
Collecting the term preceded by $\mtr{A}$, we obtain the system
\begin{eqnarray}
    \left(\frac{\rd b_1}{\rd \tau} - b_3 \right) \vv{u} +  \left(\frac{\rd b_2}{\rd \tau} - b_4 \right) \vv{U} & = & \vv{0}, \nonumber \\
    \left(\frac{\rd b_3}{\rd \tau} + w^2 b_1 \right) \vv{u} +  \left(\frac{\rd b_4}{\rd \tau} + w^2 b_2 \right) \vv{U} & = & \vv{0},
\end{eqnarray}
with an obvious solution
\begin{equation}\label{sol:b}
    b_1 =  b_4 = \cos{(w \tau)}, \qquad b_2 =   \frac{\sin{(w \tau)}}{w}, \qquad b_3 = - w \sin{(w \tau)},
\end{equation}
actually known from (\ref{sol:s}).
In the remaining part of (\ref{row:1}) and (\ref{row:2}) we change the independent variable using (\ref{eqm:ast}), and letting $v^\ast=t$ for brevity,
we find
\begin{eqnarray}
   \frac{4r}{\alpha} \left( \frac{\rd \mtr{A}}{\rd t} + \mtr{C} \mtr{A}\right) \left( b_1 \vv{u} + b_2 \vv{U}  \right) &=& \vv{0},  \\
   \frac{4r}{\alpha} \left( \frac{\rd \mtr{A}}{\rd t} + \mtr{C} \mtr{A}\right) \left( b_3 \vv{u} + b_4 \vv{U}  \right) &=& \vv{0},
 \end{eqnarray}
solved by the orthogonal matrix $\mtr{A}$, which represents rotation around $\vv{c}$ by an angle $(-\Omega t)$. Thus, the final solution consists of the
scalar equations (\ref{sol:s}) and the vector system
\begin{eqnarray}
    \vv{v} & = & \mtr{R}(\qq{q}) \left[ \cos{(w \tau)} \,\vv{u} + \sin{(w \tau)} \, \frac{\vv{U}}{w}\right], \nonumber \\
    \vv{V} & = & \mtr{R}(\qq{q}) \left[ -w \sin{(w \tau)} \, \vv{u} + \cos{(w \tau)} \, \vv{U}\right],
    \label{sol:v}
\end{eqnarray}
where $\qq{q}$ is defined as in (\ref{rotq}). Since the solutions for physical time $v^\ast(\tau)$ and distance $r(\tau)$ do not depend on rotation
of the reference frame, we omit them -- the readers may find them in \citet{StS:71}, \cite{LB:15} or any other KS-related text.

It is  common to select the reference frame rotation axis as $\vv{c}=\vv{e}_3$. In that case, the solution is further simplified, because
then the rotation matrix $\mtr{R}(\qq{q})$ does not influence $v_3$ and $V_3$. Thus the appropriate choice of the rotation axis and of the defining vector leads
to the simplest form of the solution, where only two degrees of freedom, namely $v_1, V_1$, and $v_2, V_2$, are affected by the rotation.

\section{Concluding remarks}

We dare to hope that pinpointing the presence of the defining vector in the Kustaanheimo-Stiefel transformation
may help in both the understanding and the efficient use of this ingenuous device. We do encourage those of the readers
who practice the use of KS variables, to choose the defining vector best suited for the problem at hand, instead of
inertially following the `for example' choice made by Kustaanheimo and Stiefel. Actually, a common habit in physics is to align the third axis
with a symmetry axis of the potential. It means that the KS3 set ($\vv{c}=\vv{e}_3$) should be widespread, which is true in
physics but not yet in celestial mechanics.
Of course (as noticed by a reviewer),
from a purely formal point of view, the adjustment of, say, KS1 to a different preferred direction $\vv{c}$ may be achieved
by means of a rotation matrix $\mtr{M} \in \mathrm{SO(4)}$, applied to the left-hand side of (\ref{KSvec}) or (\ref{KSD}). But then,
in the quaternion formalism, one should associate to $\mtr{M}$ a unit quaternion $\qq{m}$, so that
\begin{equation}\label{rev:1}
   \alpha\,  \qq{m} \qq{x} \bar{\qq{m}} = \qq{v} \qq{e}_1 \bar{\qq{v}},
\end{equation}
and, finally
\begin{equation}\label{rev:2}
   \alpha\,  \qq{x}  = \bar{\qq{m}} \qq{v} \qq{e}_1 \bar{\qq{v}} \qq{m}.
\end{equation}
The next step towards the form (\ref{KSgen}) is blocked by the lack of commutativity in the quaternion product, leaving the
relation of $\qq{m}$ to $\vv{c}$ unclear, and simplicity is lost, unless some trivial $\qq{m}$ has been considered.

Another way of using an arbitrary preferred direction is implicitly present in the theory of generalized
L-matrices worked out by \citet{Pole:2003}, yet its geometrical interpretation, analogous to the one we propose,
would need an additional effort.

Performing the canonical extension of the KS coordinates (point) transformation, we went a step further than usual, providing
the direct expression of new momenta $\qq{V}$ in terms of $\vv{x}$ and $\vv{X}$. It has revealed the dependence of the KS momenta
on the Cartesian angular momentum vector. The explicit relation between Fradkin tensor and Laplace vector, derived in this paper,
is another point of novelty, at least to our knowledge.

While working on some parts of the present study, we have been occasionally surprised by the power of the quaternion algebra
when applied to the KS formulation of \citet{DEF}. It was not our intention to contradict the part of their conclusions
that praised symbolic processors and lengthy calculations, but some proofs happened to be shorter than expected and we could not help it.

\begin{acknowledgements}
Not by chance, the text was completed and submitted in the day marking the tenth anniversary of the death of Prof. Andr\'e Deprit.
We would like to dedicate the work to the memory of this unforgettable man of science and a benevolent spirit of our Observatory.
We thank reviewers for their comments and bibliographic suggestions.
\end{acknowledgements}


\end{document}